\documentclass[prl,twocolumn,showpacs,amsmath,amssymb,superscriptaddress]{revtex4-1}
\usepackage{graphicx}
\usepackage{dcolumn}
\usepackage{bm}
\usepackage{bbm}
\usepackage{ulem}
\usepackage{color}
\usepackage{slashed}
\usepackage{hyperref}
\usepackage{dsfont}

\newcommand{\nc}{\newcommand}

\nc{\be}{\begin{equation}} \nc{\ee}{\end{equation}}
\nc{\bea}{\begin{eqnarray}} \nc{\eea}{\end{eqnarray}}
\nc{\bean}{\begin{eqnarray*}} \nc{\eean}{\end{eqnarray*}}

\begin{document}
\title{Universality Lost: Relation between quantizations of the Hall conductance and the edge
exponents in fractional quantum Hall effect}

\author{Jimmy A. Hutasoit}
\email{hutasoit@lorentz.leidenuniv.nl}
\affiliation{Department of Physics, The Pennsylvania State University, University Park, PA 16802, USA}
\affiliation{Instituut-Lorentz, Universiteit Leiden, P.O. Box 9506, 2300 RA Leiden, The Netherlands}

\begin{abstract}
We note an implication of chiral Luttinger liquid based edge state description of the fractional quantum Hall effect. By considering several examples that involve backward moving neutral modes, arising from either composite fermions with reverse flux attached or edge reconstruction, we show that non-universality of the edge exponent implies non-universality of the Hall conductance, as measured in the two-terminal conductance.

\end{abstract}
\maketitle

The most remarkable aspect of the quantum Hall effect, both integer and fractional, is the fact that the Hall conductance is quantized, taking only discrete set of values. This quantization is universal, in the sense that it does not depend on the details of electron interactions and edge potentials. In the case of four-terminal set up, such universality has been confirmed to the accuracy of one part per million for the integer quantum Hall effect. Even though the terms ``quantized'' and ``universal'' are not synonymous, in the quantum Hall effect they are deeply interconnected. Therefore, for brevity, in what follows we will occasionally use only one of these terms while implying both. 

The universal behavior of the quantized Hall conductance in quantum Hall effect is understood to be connected to the topological properties of the quantum states \cite{Kohmoto1985343,PhysRevB.31.3372} and this ushered in the study of topologically non-trivial insulating phases whose bulk excitations are gapped but edge/boundary states are gapless. The existence of gapless edge excitations in quantum Hall states can be understood using gauge argument \cite{PhysRevB.23.5632,PhysRevB.25.2185,PhysRevB.43.11025} and Wen proposed the chiral Luttinger liquid as the building block for the description of these edge states \cite{doi:10.1142/S0217979292000840}. 

One interesting implication of chiral Luttinger liquid based edge theory is that for simple edges, such as $\nu=1$ and $\nu=1/3$, the current-voltage relation of the (electron) tunneling between a Fermi liquid and the quantum Hall edge exhibits power law behavior with quantized and universal scaling exponent \cite{doi:10.1142/S0217979292000840} which, just like the universal behavior of the Hall conductance, is dictated solely by bulk topological properties \cite{Wen:advances1995} \footnote{Throughout this article, we limit ourselves to the case of incompressible states. For the story of compressible states, see for example \cite{Khveshchenko1999501,Khveshchenko:aa}.}. This is remarkable because of the one-to-one correspondence between the tunneling exponent and the scaling dimension of electron, where the latter generally depends strongly on the details of the interaction.

Unlike the case of the Hall conductance, however, the experimental measurements for fractional quantum Hall (FQH) states at $\nu=n/(2n \pm 1)$ \cite{RevModPhys.75.1449} and at $\nu =5/2$ \cite{Miller:2007fr} have not yet yielded a quantized tunnelling exponent and the results seem to suggest a strong sample dependence. This motivated several theoretical proposals for explaining this discrepancy \cite{PhysRevB.54.R14309,PhysRevLett.80.141,PhysRevB.59.15323,Mandal2001503,PhysRevLett.89.096801,PhysRevB.67.045303,PhysRevLett.94.166804,PhysRevB.75.165306,PhysRevLett.102.116801,PhysRevLett.88.056802,PhysRevB.68.125307}. In particular, it is found that the interplay between electron-electron interaction and confining potential at shorter distances can cause an instability that drives edge reconstruction and in the edge reconstructed phase, the quantum Hall state might lose some of its universal features, in particular, the tunneling exponent is non-quantized and non-universal \cite{PhysRevLett.88.056802,PhysRevB.68.125307}. Compared to the original state, the edge reconstructed state has at least an additional anti-parallel edge modes and as we shall see, the interaction between counter propagating modes is a necessary condition for a non-universal tunnelling exponent.

Tunneling exponent, however, is not the only observable that might lose universality due to interaction between counter propagating modes. As noted in Ref. \onlinecite{PhysRevLett.72.4129}, the interaction between counter propagating modes renders the Hall conductance non-quantized and non-universal. Even though the loss of universality in both Hall conductance and tunneling exponent have been known and studied for a while, as far as we know, the direct relationship between them has not been discussed in the literature. In this article, we aim to fill that hole. More precisely, by considering several examples of quantum Hall states with counter-propagating modes, such as those arising from composite fermions with reverse flux attachment and edge reconstructed states, we show that a quantization of the Hall conductance, as measured in the two-terminal set up, implies a quantization of the edge exponent. In other words, within the context of chiral Luttinger liquid, non-universality of the edge exponent implies non-universality of the two-terminal conductance.

Let us start by first summarizing some formulas that will be used in what follows. For their derivation, see Ref. \onlinecite{PhysRevB.57.10138}. Let us consider an edge theory whose bosonic sector is described by 
\bea
S_{b} = \frac{1}{4 \pi} \int d\tau\,dx \left(K_{ij} \, \partial_{\tau} \phi_i \, \partial_x \phi_j + V_{ij}  \, \partial_{x} \phi_i \, \partial_x \phi_j \right), \label{eq:action}
\eea
where $i, j = 1, \cdots, n$; $n$ is the number of edge modes; $K$ is a symmetric integer matrix; and $V$ is a symmetric positive matrix. The filling factor is given by $\nu = t^T \cdot K^{-1} \cdot t$,
where the vector $t$ specifies the charges of quasiparticles. As such, $K$ and $t$ are determined (modulo basis transformation) by the bulk topological properties, while $V$ parametrizes the interaction and edge potential (here, we only consider contact interaction). We say that an observable is not quantized if one can continuously tune its value by tuning $V$ and furthermore, a strong dependence on $V$ renders an observable non-universal. We note that Ref. \onlinecite{PhysRevB.57.10138} also included disorder induced tunneling terms in the action. Such terms cause a regime of parameter space to be a renormalization group (RG) attractor. It turns out that this regime is only a subspace of the parameter regime we are interested in and therefore, our result holds not only when the edge is clean but also when it is disordered. 

Continuing with our formalism, for an operator that is expressed by ${\cal O}_{\ell} = e^{i \ell_i \phi_i}$, the charge is given by $q_{\ell} = t^T \cdot K^{-1} \cdot \ell$ and its exchange statistics with respect to another operator ${\cal O}_k$ (which can be itself) is given by $\theta_{k \ell} = \pi \, k^T \cdot K^{-1} \cdot \ell$. For electron operators, the charge must be equal to unity while the exchange statistics must be that of a fermion.

In order to determine the Hall conductivity and the tunneling exponent, we need to diagonalize the action in Eq. (\ref{eq:action}). First, let us consider a basis transformation $\phi' = M_1^{-1} \cdot \phi$, under which 
\be
K' = M_1^T \cdot K \cdot M_1 =    \begin{pmatrix} 
      -\mathds{1}_{n_-} & 0 \\
      0 & \mathds{1}_{n_+}  \\
   \end{pmatrix},
\ee
where $\mathds{1}_{n_{\pm}}$ is an $n_{\pm} \times n_\pm$ identity matrix and $n_- + n_+ = n$. Next, we can diagonalize $V'=M_1^T \cdot V \cdot M_1 $ by
\be
V'' = M_2^T \cdot M_1^T \cdot V \cdot M_1 \cdot M_2,
\ee
where $V''$ is a diagonal matrix and $M_2 \in SO(n_-,n_+)$ such that $K'' = K'$. We can express the second basis transformation as $M_2 = B \cdot R$, where $R$ is an orthogonal matrix, \textit{i.e.}, the rotation, and $B$ is a positive matrix, \textit{i.e.}, the pure boost of Lorentz group. It turns out that the scaling dimension of an operator ${\cal O}_{\ell''}$ is given by 
\be
\Delta_{\ell''} = {\ell''}^T \cdot \Delta \cdot {\ell''}.
\ee
where
\be
\Delta = \frac{B^2}{2} . \label{eq:delta}
\ee 
We are particularly interested in the smallest scaling dimension of electron operators $\Delta_{\rm el}$ due to the fact that under the assumption that the outside electron couples to all the edge modes with equal strength, the scaling exponent of electron tunneling into the edge at long time scale will be given by $2 \Delta_{\rm el}$. Furthermore, the two-terminal Hall conductance is given by
\be
\sigma_H = 2 \, {t''}^T \cdot \Delta  \cdot {t''}. \label{eq:Hallc}
\ee
Here, the two-terminal conductance is defined following Refs. \onlinecite{PhysRevB.23.6851,PhysRevLett.68.1220}, where one applies electric field along the edge and evaluate the current response.

We would like to note that the parameters of the boost $B$ describe the mixing between counter-propagating modes, while the parameters of the rotation $R$ describe the mixing between modes propagating along the same direction. Since Eq. (\ref{eq:delta}) shows that the non-trivial part of $\Delta$ only depends on $B$ (but not $R$), the renormalization, and thus the non-universality, of the Hall conductance and scaling dimensions of operators depend on the mixing between counter-propagating modes.

Now we are ready to consider some examples of FQH states that features backward moving neutral modes. First, let us treat the case of FQH states arising from composite fermions with reverse flux attachment. The state with filling factor $\nu = \frac{n}{2 p n -1}$ is described by
\be
K = -\mathds{1}_n + 2 p \, C_n, \qquad t = (1,\cdots,1)^T,
\ee
where $C_n$ is an $n\times n$ matrix whose entries are all equal to $1$. In a basis where the $K$-matrix is diagonal, we have
\be
K = {\rm diag}(2p n -1, -1, \cdots, -1), \quad t= (\sqrt{n},0, \cdots, 0)^T. \label{eq:Ktforreverseflux}
\ee
In this basis, we have a forward moving charge mode and $n-1$ backward moving neutral modes. In general, these modes are not the eigenmodes as we expect interaction to mix them. 

Parametrizing the boost such that
\be
B^2 = \begin{pmatrix} 
      \gamma & \beta_1 \gamma & \beta_2 \gamma & \cdots & \beta_{n-1} \gamma \\
      \beta_1 \gamma & 1+ \frac{\beta_1^2 \gamma^2}{\gamma+1} & \frac{\beta_1 \beta_2 \gamma^2}{\gamma+1} & \cdots &  \frac{\beta_1 \beta_{n-1} \gamma^2}{\gamma+1}\\
      \beta_2 \gamma & \frac{\beta_1 \beta_2 \gamma^2}{\gamma+1} & 1+ \frac{\beta_2^2 \gamma^2}{\gamma+1}  & \cdots &  \frac{\beta_2 \beta_{n-1} \gamma^2}{\gamma+1} \\
      \vdots & \vdots & \vdots & & \vdots\\
      \beta_{n-1} \gamma & \frac{\beta_{n-1} \beta_1 \gamma^2}{\gamma+1} &  \frac{\beta_{n-1} \beta_2 \gamma^2}{\gamma+1} & \cdots & 1+ \frac{\beta_{n-1}^2 \gamma^2}{\gamma+1}
      \end{pmatrix}, \label{eq:boost}
\ee
where $\gamma = 1/\sqrt{1-\beta^2}$, $\beta^2= \sum_{i=1}^{n-1} \beta_i^2$ and $|\beta|\leq1$, yields
\be
\sigma_H = \frac{\nu}{\sqrt{1-\beta^2}}.
\ee
This means that in order for the two-terminal conductance to be quantized and taking the ``correct" value, all of the boost parameters $\beta_i$'s must vanish. In other words, since $\beta_i$'s describe the mixing between the charged mode and the counter-propagating neutral modes, Hall conductance is quantized if and only if the charged mode is decoupled from all the backward moving neutral modes. In this case, however, $B^2$ is just an identity matrix, and therefore, the scaling dimension of the electron operator will also be quantized and universal.

For the next case, let us consider edge reconstructed Laughlin states and edge reconstructed Pfaffian. For Laughlin state and the bosonic sector of Pfaffian state, the edge reconstructed state is described by 
\be
K =    \begin{pmatrix} 
      -m & 0 & 0 \\
      0 & m & 0 \\
      0 & 0 & m
   \end{pmatrix},
\qquad t =    \begin{pmatrix} 
      1  \\
      1  \\1  \\
   \end{pmatrix}, \label{eq:edgerecK}
\ee
where $m$ is an odd integer for Laughlin state and $m=2$ for Pfaffian state. Doing a basis transformation such that $K \rightarrow W \cdot K \cdot W^T$, with
\be
W = \begin{pmatrix} 
      \frac{\sqrt{2}}{\sqrt{m}} & -\frac{1}{\sqrt{2m}} & -\frac{1}{\sqrt{2m}}  \\
      0 & \frac{1}{\sqrt{2m}}  & -\frac{1}{\sqrt{2m}}   \\
      -\frac{1}{\sqrt{m}} & \frac{1}{\sqrt{m}}  & \frac{1}{\sqrt{m}} 
   \end{pmatrix},
\ee
we obtain
\be
K =    \begin{pmatrix} 
      -1 & 0 & 0 \\
      0 & 1 & 0 \\
      0 & 0 & 1
   \end{pmatrix},
\qquad t =    \begin{pmatrix} 
      0  \\
      0  \\\frac{1}{\sqrt{m}}  \\
   \end{pmatrix}.
\ee
In this basis, we have a forward moving charge mode and a couple of anti-parallel neutral modes. As before, these modes are generally not the eigenmodes as we expect interaction to mix them. 

Parametrizing the boost exactly as in Eq. (\ref{eq:boost}) but with $n=3$ yields
\be
\sigma_H = \frac{1}{m} \left(1+ \frac{\beta_2^2}{1-\beta^2 + \sqrt{1-\beta^2}}\right).
\ee
This means that in order for the Hall conductance to be quantized at the correct value, $\beta_2$ must vanish. Even though the quantization of the Hall conductance requires the charged mode to be decoupled from the backward moving neutral move, the two anti-parallel neutral modes can still interact. Nevertheless, as we shall see, this interaction does not render the smallest scaling dimension of the electron operators to be non-universal.

The electron operators can be written as
\be
{\cal O}_{\rm el} = \exp[i (x \phi_{n1} + y \phi_{n2} + \sqrt{m} \phi_c)],
\ee
where $\phi_c$ is the charged mode, $\phi_{n1,n2}$ are the backward and forward moving neutral modes, respectively, and $y^2- x^2 = 2 p$, where $p$ is an integer. This condition needs to be satisfied in order for the electron operators to have fermionic statistics. If Hall conductance is quantized, the scaling dimension of the electron operator is then given by
\be
\Delta_{\rm el} = \frac{x^2 + 2 \beta_1 x y + y^2}{2\sqrt{1-\beta_1^2}} + \frac{m}{2}.
\ee
(For Pfaffian, this is only the bosonic part of the electron operator and the full operator is obtained by multiplying this expression with the Majorana fermion.) It is then easy to see that the long time behavior of electron tunneling will be dominated by the electron operator with scaling dimension $\Delta_{\rm el} = m/2$. To see that, we note that $1 \geq \beta_1\geq -1$ and thus, $x^2 + 2 \beta_1 xy+y^2 \geq |x|^2 -2 |x| |y| + |y|^2 = \left(|x|-|y|\right)^2 \geq 0$ where the minimum can always be reached by setting $x=y=0$ regardless of the value of $\beta_1$.

Therefore, when the Hall conductance is quantized, then the scaling dimension of the most dominant electron operator is also quantized to be $\Delta_{\rm el} = m/2$ for edge reconstructed Laughlin state (\textit{c.f.}, \cite{PhysRevB.49.8227}) and  $\Delta_{\rm el} = 3/2$ for Pfaffian (\textit{c.f.}, \cite{PhysRevB.90.165104}). In the light of tunneling experiments such as that of Refs. \onlinecite{RevModPhys.75.1449,Miller:2007fr}, where the edge exponent is found to be non-universal (while the Hall conductance is quantized), edge reconstruction has been proposed as a mechanism that results in the non-universal behavior of the edge \cite{PhysRevLett.88.056802,PhysRevB.68.125307}. However, our result clearly shows that edge reconstruction as described by Eq. (\ref{eq:edgerecK}) \textit{cannot} be the explanation of the non-universal behavior found in tunneling experiments.

As the last examples, let us consider other FQH states with ${\rm dim} [K] = 3$ and anti-parallel neutral modes, such as $\nu = 1 \pm \frac{2}{4 p -1}$. As before, we can do a basis transformation such that
\be
K =    \begin{pmatrix} 
      -1 & 0 & 0 \\
      0 & 1 & 0 \\
      0 & 0 & 1
   \end{pmatrix},
\qquad t =    \begin{pmatrix} 
      0  \\
      0  \\\sqrt{\nu}  \\
   \end{pmatrix}. \label{eq:Ktgeneral}
\ee
Using the same parametrization for the boost as above, we see that in order for the two-terminal conductance to be quantized $\beta_2$ must vanish. Furthermore, the scaling dimension of the electron operator is
\be
\Delta_{\rm el} = \frac{x^2 + 2 \beta_1 x y + y^2}{2\sqrt{1-\beta_1^2}} + \frac{1}{2\nu}, \label{eq:elscaling5/3}
\ee
but with the condition 
\be
y^2- x^2 = 2 p+1- \frac{1}{\nu}, \label{eq:fermicond}
\ee
where again, $p$ is an integer. In this case, the first term of Eq. (\ref{eq:elscaling5/3}) is positive definite because $x=y=0$ is not a solution to Eq. (\ref{eq:fermicond}). Solving Eq. (\ref{eq:fermicond}) for $y$, substituting the solution into Eq. (\ref{eq:elscaling5/3}) and then minimizing it with respect to $x$, we obtain
\be
\Delta_{\rm el}^{\rm min} = \left|p_{\rm min} + \frac{1}{2} - \frac{1}{2\nu}\right|+ \frac{1}{2\nu}, \label{eq:elscalingmin}
\ee
where $p_{\rm min}$ is an integer chosen to minimize the first term. Since all dependence on $\beta_1$ has dropped off the smallest scaling dimension of the electron operators, we again conclude that if the two terminal conductance is quantized then the electron tunneling exponent will also be quantized. 


Some discussions are in order. In this article, we have considered three classes of FQH states: $\nu=n/(2n \pm 1)$, edge reconstructed $\nu=1/m$ and $\nu = 1 \pm \frac{2}{4 p -1}$; all of which contain counter-propagating modes. We started by showing that the decoupling between the forward moving charged mode and the backward moving neutral modes is the sufficient and necessary condition for quantized Hall conductance, as measured in two-terminal set up. Since the parameter space in which such decoupling occurs is a lot smaller than the whole parameter space, this begs the question of what mechanism confines us to the subspace of parameter space in which the forward moving charged mode and the backward moving neutral modes are decoupled. One such mechanism was introduced in Ref. \onlinecite{PhysRevLett.72.4129}, where it was shown that edge disorder can restore the quantization of Hall conductance because in the presence of disorder, there is an RG fixed point, the so-called Kane-Fisher-Polchinski (KFP) fixed point, at which the Hall conductance takes the correct quantized value. This KFP fixed point is obviously a subspace of the parameter subspace in which the forward moving charged mode and the backward moving neutral modes are decoupled. 

Anticipating the possibility of other mechanisms that can restore the quantization of Hall conductance, in this article we did not make any assumptions of what such mechanisms should be. Instead of limiting ourselves to a subspace of the parameter subspace in which the decoupling between the forward moving charged mode and the backward moving neutral modes occurs, we simply observed that as long as the forward moving charged mode and the backward moving neutral modes are decoupled, the tunneling exponent is universal. Providing a mechanism that will confine us to a subspace of the parameter subspace we considered above, such as by introducing disorder, obviously will not change the result. For the particular case of disordered edge, in a sense, what we did can be thought of as a generalization of Refs. \onlinecite{PhysRevB.57.10138,PhysRevB.49.8227,PhysRevB.90.165104}, where the authors studied the tunneling exponents at the KFP fixed points of the three classes of FQH states we considered here. 

Taking into account the two statements: 
\begin{itemize}
\item the decoupling between the forward moving charged mode and the backward moving neutral modes is the sufficient condition for universal tunneling exponent, 
\item this modes decoupling is the sufficient and necessary condition for universal Hall conductance, 
\end{itemize} 
we concluded that quantization of the Hall conductance, as measured in two-terminal set up, implies the quantization of tunneling exponent. Equivalently, at least within the framework of chiral Luttinger liquid theory, a non-universal tunneling exponent implies a non-universal Hall conductance. 

Lastly, let us comment shortly on the case of four-terminal conductance. In this case, even though we do not have a somewhat general formula akin to Eq. (\ref{eq:Hallc}), at least for $\nu=2/3$, the decoupling between the charged mode and the backward moving neutral mode is also the sufficient and necessary condition for quantized and universal four-terminal conductance \cite{PhysRevLett.72.4129}. Therefore, in that case, a non-universal tunneling exponent also implies a non-universal four-terminal conductance.

\section*{Acknowledgement}
I would like to thank Jainendra K. Jain, Joel Moore and Diptiman Sen for insightful discussions. The early part of this work is supported by NSF grant DMR-1005536 and DMR-0820404 (Penn State MRSEC), and later by the Netherlands Organization for Scientific Research (NWO/OCW) through the D-ITP consortium.

\bibliography{References}
\end{document}